\documentclass[fleqn,12pt,twoside]{article}
\usepackage{espcrc1}

\input{epsf.tex}
\newcommand{\AmS}{{\protect\the\textfont2
  A\kern-.1667em\lower.5ex\hbox{M}\kern-.125emS}}

\title{Collective clusterization effects in light heavy ion reactions}

\author{Raj K. Gupta\address[CHD]{Physics Department, Panjab University, 
        Chandigarh-160014, India},      
        M. Balasubramaniam\addressmark,
        Rajesh Kumar\addressmark,
        Dalip Singh\addressmark,
        and 
        C. Beck\address{Institut de Recherches Subatomiques, UMR7500, IN2P3/ Universit\'e 
        Louis Pasteur, \\
        B.P. 28, F-67037 Strasbourg Cedex 2, France}
        }
       
\begin{document}

% typeset front matter
\maketitle

\begin{abstract}
The collective clusterization process, proposed for intermediate mass fragments 
(IMFs, 4$<$A$\le$28, 2$<$Z$\le$14) emitted from the hot and rotating compound nuclei formed 
in low energy reactions, is extended further to include also the emission of light particles 
(LPs, A$\le$4, Z$\le$2) from the fusion-evaporation residues. Both the LPs and IMFs are 
treated as the dynamical collective mass motion of preformed clusters through the 
barrier. Compared to IMFs, LPs are shown to have different characteristics, and the 
predictions of our, so-called, dynamical cluster-decay model are similar to those 
of the statistical fission model. 
\end{abstract}

\section{INTRODUCTION}

For A$_{CN}^*$$\ge$48, the compound nucleus (CN) so formed decays subsequently by the emission 
of mainly the LPs (n, p, $\alpha$ and $\gamma$-rays), and a significant (5-10\%) decay 
strength is observed for the complex IMFs. The measured angular distributions and energy 
spectra for IMFs are consistent with fission-like decays of the respective CN. 

The statistical Hauser Feshbach (HF) analysis, given for  LPs emission, is also used to 
include the IMFs emission (BUSCO code \cite{campo91}, or the Extended HF scission-point model 
(EHFM) \cite{matsuse97}). For IMFs, a more accepted process is the binary fission of the CN 
(fusion-fission), worked out in the statistical scission-point fission model \cite{charity88} 
or the saddle-point "transition-state" model (TSM) \cite{sanders89,sanders91}, with the LPs 
still treated within the HF method. 

The recently advanced dynamical cluster-decay model (DCM) \cite{gupta02} include the missing 
structure information of the statistical fission models, by taking each fragment (LP or IMF) 
preformed in the CN, which subsequently follows the dynamical cluster-decay process rather 
than the fission process of the statistical model. The structure information influences the 
observed yields strongly through the observed strong resonance behaviour in the measured 
excitation functions of large-angle elastic and inelastic scattering yields (see, e.g., 
\cite{beck01}), which enters the DCM via the preformation probability of the fragments. 

The DCM for hot and rotating compound systems is a reformulation of the preformed cluster 
model (PCM) of Gupta et al. \cite{malik89,kumar94} for ground-state decays, briefly developed 
in Section 2. The application of DCM to $^{32}$S+$^{24}$Mg$\rightarrow$$^{56}$Ni$^*$ is 
discussed in section 3. The $^{56}$Ni$^*$ has a negative Q-value (Q$_{out}$), with experimental 
mass distributions of both LPs and IMFs being available \cite{sanders89}. The model has also 
been applied successfully for the positive 
Q-value $^{116}Ba^*$ system.
A summary of our 
results is presented in section 4.

\section{THE DCM FOR HOT AND ROTATING NUCLEI}

The DCM uses collective coordinates of mass (charge) asymmetry 
$\eta$={{(A$_1$-A$_2$)}/{(A$_1$+A$_2$)}} ($\eta _Z$={{(Z$_1$-Z$_2$)}/{(Z$_1$+Z$_2$)}}), 
and relative separation R, characterizing, respectively, the nucleon-exchange between the 
outgoing fragments, and the incident channel kinetic energy (E$_{cm}$) transferred to internal 
excitation of outgoing channel (E$_{CN}^*$+Q$_{out}$(T)=TKE(T)+TXE(T); TXE and TKE as 
the total excitation and kinetic energy), since R$\equiv$R(T,$\eta$). For 
decoupled R-,$\eta$-motions, in terms of partial waves, the fragment 
formation or CN decay cross-section 
%\cite{gupta02} 
\begin{equation}
\sigma={\pi \over k^2}\sum_{l=0}^{l_c}(2l+1)P_0P; \qquad k=\sqrt{2\mu E_{cm}\over {\hbar^2}}.
\label{eq:1}
\end{equation}
The preformation probability P$_0$, refering to $\eta$-motion, is the solution of stationary 
Schr\"od- inger eqn. in $\eta$ (with T-dependent collective fragmentation potentials
\cite{davidson94,myers66,royer92}), at a fixed 
\begin{equation}
R=R_a=C_t(\eta,T)+\overline{\Delta R}(T).
\label{eq:2}
\end{equation}
\begin{figure}[htb]
\vspace*{-1.4cm}
\centerline{\epsfxsize=6.8in\epsffile{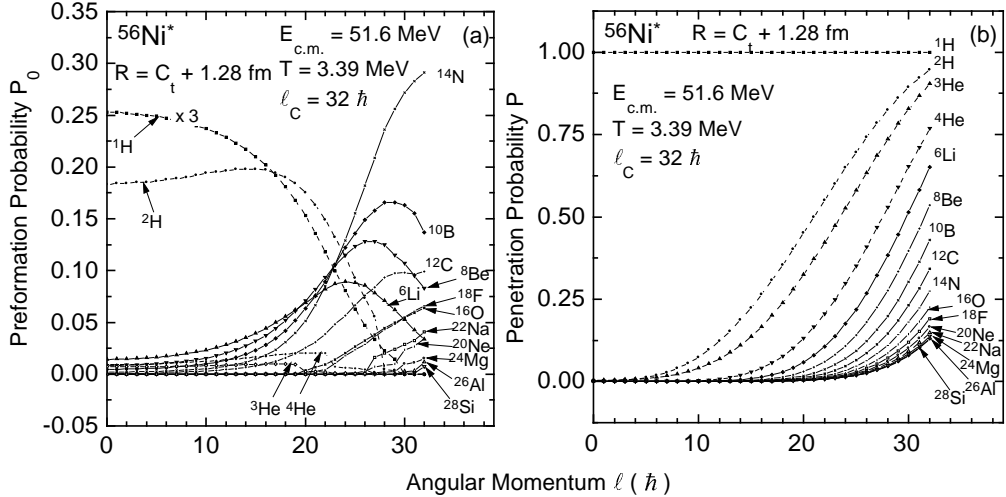}}
\vspace*{-15.1cm}
\caption{(a) $P_0(\ell)$ (b) $P(\ell)$ for decay of $^{56}$Ni$^*$ at T=3.39 MeV.
}
\label{Fig.1}
\end{figure}

\vspace*{-0.45cm}
\par\noindent
C$_i$ are S\"usmann radii and T is from
E$_{CN}^*$=E$_{cm}$+Q$_{in}$=$\left ({A/9}\right )${T}$^2$-T.
$\overline{\Delta R}$(T), the only parameter of model, simulating the two-centre 
nuclear shape, is similar to neck-length used in 
statistical models \cite{matsuse97,sanders91}. The penetrability P, refering to R-motion, 
is the WKB integral, with V(R$_a$)=V(R$_b$)=Q$_{eff}$. R$_b$ is  
second turning point. Q$_{eff}$, the effective Q-value or TKE(T,$\ell$=0), is
Q$_{eff}$(T)=B(T)-[B$_1$(T=0)+B$_2$(T=0)]=TKE(T)=V(R$_a$),
for the CN at T to decay to two observed cold fragments (T=0).  
This transition occurs by emitting LPs of energy 
E$_x$=B(T)-B(0)=Q$_{eff}$(T)-Q$_{out}$(T=0). Instead, here we use Eq. (\ref{eq:2}). 
%The critical $\ell$-value, 
$\ell _c=R_a\sqrt{2\mu [E_{cm}-V(R_a,\eta _{in},\ell=0)]}/\hbar$,
with $\mu$ and R$_a$ referring to entrance channel $\eta _{in}$.
\begin{figure}[htb]
\vspace*{-0.8cm}
\centerline{\epsfxsize=7.08in\epsffile{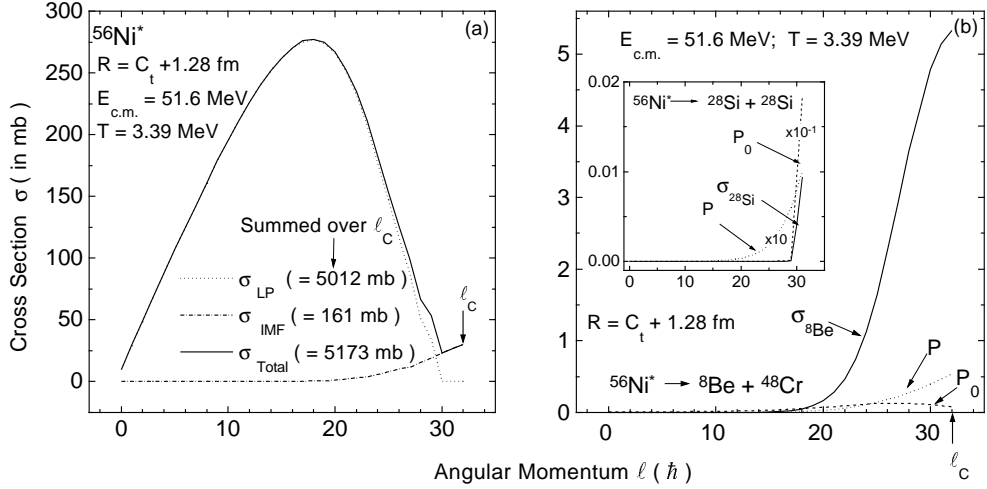}}
\vspace*{-17.25cm}
\caption{The calculated (a) $\sigma_{LP}(\ell)$, $\sigma_{IMF}(\ell)$ and their sum 
$\sigma_{Total}(\ell)$; (b) $P_0(\ell)$, $P(\ell)$ and $\sigma_{IMF}(\ell)$ for $^8$Be
and $^{28}$Si decays of $^{56}$Ni$^*$ at T=3.39 MeV.
}
\label{Fig.2}
\end{figure}

\vspace*{-1.0cm}
\section{CALCULATIONS}

Fig. 1 gives the variation of $P_0$ and $P$ with $\ell$ for both the LPs ($A\le 4$) and 
IMFs (A$>4$) refering to energetically favoured fragments.
Interestingly, the behaviour of LPs is different from that of the IMFs; 
whereas for LPs $P_0$ decreases with increase of $\ell$, for IMFs it 
increases as $\ell$ increases and then starts to decrease at a large $\ell$-value. 
Thus, LPs contribute to $P_0$ for smaller $\ell$-values 
whereas the same for IMFs is important for higher $\ell$'s ($>18 \hbar$). 
Similarly, the $P$'s for LPs are large at all $\ell$-values, whereas the same for 
$\ell \le 18 \hbar$ is nearly zero for all IMFs. This means that the lower $\ell$'s 
contribute only towards LPs cross-section $\sigma_{LP}$ and the higher $\ell$'s ($>18 \hbar$) 
to fission-like, IMFs cross-section $\sigma_{IMF}$, as shown in Fig. 2(a). The total 
cross-section 
$\sigma_{Total}=\sigma_{LP}+\sigma_{IMF}$
is also plotted, and the summed up cross-sections are given in braces, to be
compared with experiments. 
 
Fig. 2(b) illustrates the contributions of individual IMFs. Apparently, the lighter 
IMFs contribute more than the heavier IMFs towards $\sigma_{IMF}$, pointing to
the observed favourable asymmetric mass distribution. Interestingly, similar
results, as presented by Figs. 2(a) and 2(b), are obtained by the statistical fission 
model \cite{matsuse97,sanders89} (see e.g. Fig. 14 in \cite{sanders89}). 

Fig. 3(a) shows the summed up $\sigma$ for decay of $^{56}$Ni$^*$ at $E_{c.m.}=$51.6 MeV 
to both the LPs and even-A, N=Z IMFs, along with the TSM and EHFM calculations, compared 
with the experimental data \cite{sanders89}. The TSM and EHFM 
for LPs are the HF results, whereas the DCM treats both the LPs and IMFs emissions on similar 
footings. For the LPs emission, the measured $\sigma_{LP}=1050\pm 100$ mb but the 
separate yield for each emitted light particle is not available for a direct comparison. 
First of all, we notice in Fig. 3(a) that the unphysical discontinuity at the point between 
$A_2=$4 and 6 in TSM is not present in our DCM results. 
In DCM, $\sigma_{LP}$ is over-estimated by a large factor of $\sim$5, a discrepancy that
gets removed if mass-one-particle (proton) is replaced with neutron, for example.
This signifies the importance of the contributing particles to $\sigma_{LP}$. 
For the IMFs, the general comparison between the 
data and DCM is of similar quality as for the TSM or EHFM model, at least for $A_2\le 22$. 
The use of $Q_{eff}$ improves the comparison slightly. 

Fig. 3(b) shows the average 
$\overline{TKE}={\sum_{\ell=0}^{\ell_c}{\sigma_{\ell}}(TKE)_{\ell}/\sigma}$, with
$\sigma=\sum\sigma_{\ell}$, compared with experimental data \cite{sanders89}.  
Apparently, the comparisons are reasonably good. The maximum $\ell$-value is less than 
$\ell_c$-value, which may be due to the use of $\overline{\Delta R}$ instead of actual 
${\Delta R}(\eta)$. 
\begin{figure}[htb]
\vspace*{-1.25cm}
\centerline{\epsfxsize=6.5in\epsffile{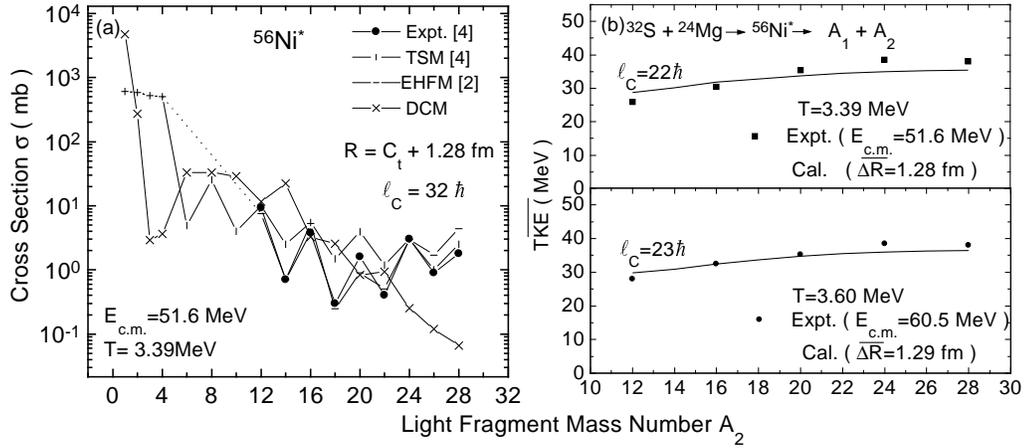}}
\vspace*{-15.2cm}
\caption{(a) The calculated $\sigma (A_2)$ on DCM, EHFM \cite{matsuse97} 
and TSM \cite{sanders89}, compared with  experimental data  \cite{sanders89}; 
(b) The measured and calculated $\overline{TKE}$ at E$_{c.m.}$=51.6, 60.5 MeV.
}
\label{Fig.3}
\end{figure}

\vspace*{-0.8cm}
\section{CONCLUSIONS}

The dynamical collective clusterization is shown as an alternative good decay
process for both the LPs and IMFs produced in low energy reactions. 
The IMFs are shown to be produced as clusters, like in cluster radioactivity, but 
from excited CN. The LPs are also 
considered to be of the same origin, without invoking any statistical evaporation process.

\end{document}